\documentclass[aps,prl,reprint, superscriptaddress,amsmath,amssymb,showpacs]{revtex4-1}

\usepackage{graphicx}
\begin{document}


\title{Coupling of a Single Nitrogen-Vacancy center in diamond to a Fiber-Based Microcavity}



\author{Roland Albrecht}
\author{Alexander Bommer}
\affiliation{Universit\"at des Saarlandes, Fachrichtung 7.2 (Experimentalphysik), Campus E2.6, 66123 Saarbr\"ucken, Germany}
\author{Christian Deutsch}
\affiliation{Laboratoire Kastler Brossel, ENS/UPMC-Paris 6/CNRS, 24 rue Lhomond, 75005 Paris, France}
\affiliation{Menlo Systems GmbH, 82152 Martinsried, Germany}
\author{Jakob Reichel}
\affiliation{Laboratoire Kastler Brossel, ENS/UPMC-Paris 6/CNRS, 24 rue Lhomond, 75005 Paris, France}
\author{Christoph Becher}
\email[]{christoph.becher@physik.uni-saarland.de}
\affiliation{Universit\"at des Saarlandes, Fachrichtung 7.2 (Experimentalphysik), Campus E2.6, 66123 Saarbr\"ucken, Germany}

\date{\today}

\begin{abstract}
We report on the coupling of a single nitrogen-vacancy (NV) center in a nanodiamond to a fiber-based microcavity at room temperature. Investigating the very same NV center inside the cavity and in free space allows us to systematically explore a regime of phonon-assisted cavity feeding. Making use of the NV center's strongly broadened emission, we realize a widely tunable, narrow band single photon source. A master equation model well reproduces our experimental results and predicts a transition into a Purcell-enhanced emission regime at low temperatures.
\end{abstract}

\pacs{42.50.Pq, 42.50.Ct, 42.81.Qb, 61.72.jn}

\maketitle


The coupling of single color centers in diamond, such as nitrogen-vacancy(NV)~\cite{Kurtsiefer00} or silicon-vacancy (SiV)~\cite{Neu11} centers, to an optical microcavity is considered an essential building block for applications in quantum information~\cite{Prawer08, Aharonovich11, Benjamin09} or magnetometry~\cite{Acosta10}. For NV centers, the cavity coupling can be employed for an efficient readout of the spin state, e.g. via enhanced spontaneous emission~\cite{Su08-2}, enhanced absorption~\cite{Acosta10} or state-dependent cavity reflection~\cite{Young09}. Recent years have witnessed a large progress in coupling color centers to microcavities, e.g. in hybrid approaches combining diamond nanocrystals and nondiamond cavities (for a review see~\cite{Aharonovich11}) or integrated all-diamond schemes (e.g.~\cite{RiedrichMoeller12, Faraon12}).

We here demonstrate a new hybrid approach by coupling a single NV center located in a diamond nanocrystal to a fiber-based Fabry-Perot cavity~\cite{Colombe07, Hunger10} operated at room temperature. Such a cavity type has recently been employed for coupling to quantum dots~\cite{Muller09, Sanchez12} and ensembles of NV centers~\cite{Kaupp13}. The cavity consists of a plane dielectric mirror onto which \hyphenation{na-no-dia-monds}nanodiamonds (NDs) are spin-coated and an optical fiber facet which has been, prior to deposition of a dielectric multilayer coating, laser machined to yield a concave imprint~\cite{Hunger10}. This open-access setup has several advantages: the emission is automatically fiber coupled, the cavity can be easily tuned and the emitter-cavity coupling can be continuously varied.

Usually, the NV center's broad emission bandwidth spanning over $100$~nm is considered a drawback for coupling to narrow cavity modes: in such a case one does not expect a Purcell enhancement of spontaneous emission as only a fraction $\kappa/\gamma^\star$ of the emission couples to the cavity mode. The Purcell factor then can be approximated as $F \approx [4g^2/(\kappa \gamma)] \kappa/\gamma^\star = 4g^2/(\gamma\gamma^\star)$ where $g$ is the emitter-cavity coupling rate, $\kappa$ the cavity loss rate, $\gamma$ the emitter's natural linewidth and $\gamma^\star$ the homogeneously broadened linewidth. For the NV center at room temperature $\gamma^\star$ typically is many orders of magnitude larger than $g$, thus $F \ll 1$. As a consequence the cavity would only exert a spatial and spectral filtering on the NV center emission.

Nevertheless, we show here that one can use the strong broadening as a resource for cavity coupling~\cite{Auffeves09} beyond simple filtering effects. As the NV emission lines (zero phonon line (ZPL) and phonon side-bands (PSBs)) at room temperature are predominantly broadened by phonon scattering, we here explore a regime of emission feeding from a broad, off-resonant emitter into a narrow cavity mode mediated by pure dephasing~\cite{Auffeves09}. In this regime the cavity does not alter the emitter's lifetime, but the emission is efficiently channeled into the cavity mode. This phonon-assisted coupling regime has been explored theoretically~\cite{Auffeves09, Auffeves10, Wilson-Rae02} and investigated experimentally~\cite{Hohenester09, Ates09, Hennessy07, Press07} for the case of semiconductor quantum dots. However, as opposed to many solid state systems, we can here investigate the very same individual NV center's emission both into free space and cavity coupled. Our system therefore is an ideal test bench to study the phonon-assisted coupling regime. 

\begin{figure}[h!!!tbp]
\centering\includegraphics[width=\columnwidth]{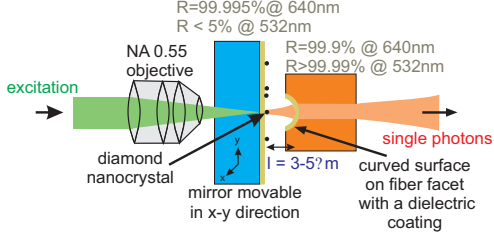}
\caption{Cavity Setup: Fabry-Perot cavity consisting of a concave mirror on a fiber facet and a plane mirror onto which NDs have been spin coated. For further details see text.}
\label{setup}
\end{figure}

The cavity setup is shown in Fig.~\ref{setup}. The coating on the plane mirror is designed to yield a reflectivity $R=99.995$\% at the NV ZPL wavelength ($\approx 640$~nm) and less than $5$\% at $532$~nm. Further, it is optimized such that the field amplitude is maximal at the ND position. The coating on the fiber is designed to provide $R=99.9$\% at $640$~nm and $R=99.99$\% at $532$~nm such that the dominant output channel for NV photons is into the fiber. The measured finesse of the cavity is up to $\mathcal{F}=3500$ at $640$~nm. 
For wavelengths $>680$~nm the finesse decreases rapidly due to increased transmissions of the coatings. The imprint on the facet has a radius of curvature of $71.6$~$\mu$m, a diameter of $20.6$~$\mu$m, a maximal depth of $1.9$~$\mu$m and a surface roughness $< 0.2$~nm. Onto the plane mirror a solution containing NDs (diameter $< 100$~nm) is spin coated. Because of their size and occasional agglomeration NDs in the cavity induce additional scattering losses leading to a reduced finesse (for details see Supplemental Material~\cite{Supp}).

The mirror is mounted on a translation stage equipped with piezo-stepper motors allowing for lateral positioning of the mirror. The fiber is glued onto a home-built piezo-driven flexure mount to control the cavity length. The NV center is excited with a laser at $532$~nm through the plane mirror using a microscope objective ($100\times$, N.A. $0.55$). The emission from the cavity is collected via the fiber and sent, after filtering out residual pump laser light, either to a Hanbury Brown-Twiss (HBT) \hyphenation{in-ter-fero-meter}interferometer or a spectrometer.

To characterize the NV center's emission into free space outside the cavity the fiber is replaced by a second microscope objective ($100\times$, N.A. $0.8$) in front of the mirror to collect the fluorescence. By laterally scanning the plane mirror we identify NDs containing single NV centers confirmed by $g^{(2)}$ measurements (see below). Figure~\ref{spectrum}(a) displays the spectrum of the single NV center showing the ZPL at $639$~nm and the PSB. To gain information on the PSB structure the spectrum is fitted with 8 Lorentzians~\cite{Supp}. To evaluate the emitter's brightness we perform a saturation measurement and obtain a saturation count rate of $I_\infty = 2.9\times 10^5$~counts/s [detected in the spectral window $650-750$~nm, corresponding to a spectral density of $\approx 4.5$~photons/(s~GHz)] and a saturation power of $P_\mathrm{sat} = 0.46$~mW.

\begin{figure}[htbp]
\centering\includegraphics[width=\columnwidth]{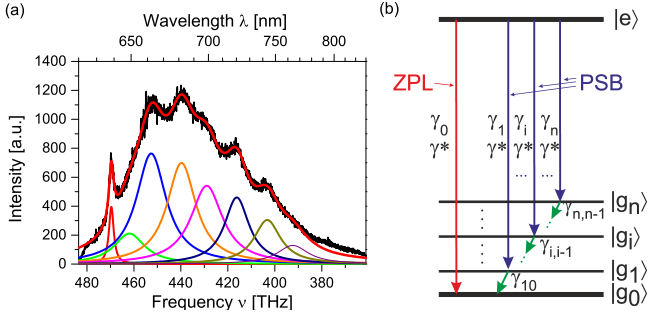}
\caption{(a) Room-temperature emission spectrum of a single NV center in a ND on a plane mirror. The spectrum has been fitted with 8 Lorentzian lines. (b) Model of the NV center with $n$ vibronic ground states $|g_i\rangle$ and one excited state $|e\rangle$. Broadening of the lines is due to spontaneous emission ($\gamma_i$), pure dephasing ($\gamma^\star$) and emission of phonons ($\gamma_{i,i-1}$).}
\label{spectrum}
\end{figure}

\begin{figure}[htbp]
\centering\includegraphics[width=\columnwidth]{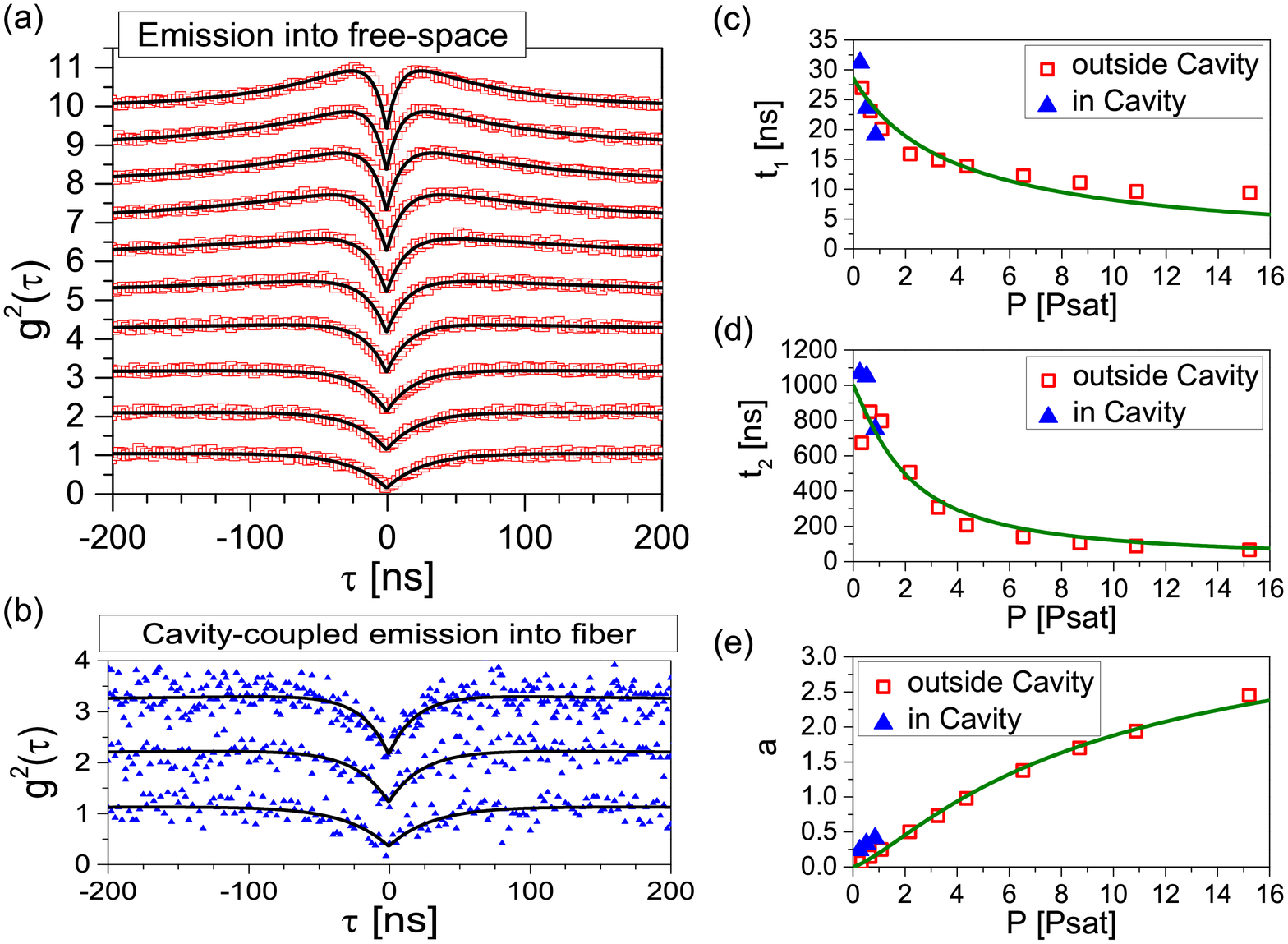}
\caption{(a) Open squares: Intensity correlation $g^{(2)}(\tau)$ measurements at different excitation powers $P$ for a single NV center on a plane mirror outside the cavity. The curves are shifted vertically by 1 each for better visibility. $P=$ $0.32\times$, $0.65\times$, $1.1\times$, $2.2\times$, $3.3\times$, $4.3\times$, $6.5\times$, $8.7\times$, $10.9\times$, $15.2\times P_\mathrm{sat}$ from bottom to top. Lines: Fit curves of $g^{(2)}(\tau) = 1 - (1+a)e^{-|\tau|/\tau_1} + a e^{-|\tau|/\tau_2}$. (b) Triangles: $g^{(2)}$ measurements at different excitation powers with the same NV center as in (a) inside the cavity. $P=$ $0.25\times$, $0.51\times$, $0.84\times P_\mathrm{sat}$ from bottom to top. Lines: Fit curves with the same function as in (a). (c) - (e) Squares and triangles: Model parameters $a$, $\tau_1$, $\tau_2$ obtained from the fits in (a) and (b). Solid line: Theoretical dependence of model parameters on excitation power.}
\label{g2meas}
\end{figure}

To verify the presence of a single NV center and to deduce its population dynamics the intensity correlation $g^{(2)}(\tau) = \langle : I(t+\tau) I(t):\rangle / \langle I(t)\rangle^2$ is measured with the HBT interferometer at different excitation powers as shown in Fig.~\ref{g2meas}(a). The measured data is fitted with $g^{(2)}(\tau) = 1 - (1+a)e^{-|\tau|/\tau_1} + a e^{-|\tau|/\tau_2}$, the form expected for a three level system~\cite{Kurtsiefer00, Neu11}. The obtained fit parameters $\tau_1$, $\tau_2$, $a$ are shown in Figs~\ref{g2meas}(c)-(e). From the limiting values of these parameters we estimate the rates of the NV center population dynamics  yielding a theoretical power dependence of the parameters $\tau_1$, $\tau_2$ and $a$~\cite{Kurtsiefer00, Neu11} as shown by the solid lines in Figs.~\ref{g2meas}(c)-(e)~\cite{Supp}.

\begin{figure}[htbp]
\centering\includegraphics[width=\columnwidth]{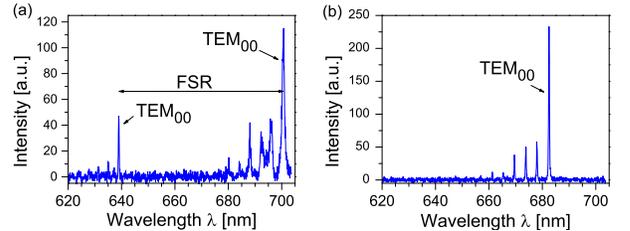}
\caption{Cavity emission spectrum at effective cavity lengths (a) $l=3.5$~$\mu$m and (b) $l=3.1$~$\mu$m. FSR denotes the free spectral range, TEM$_{00}$ the fundamental transversal modes. The smaller modes at shorter wavelengths are higher order transversal modes.}
\label{incav}
\end{figure}

After free-space characterization we investigate the same emitter (confirmed by a lateral scan of the mirror) coupled to the fiber cavity. We record the cavity emission spectra from the output of the fiber for two different cavity lengths of $l=3.5$~$\mu$m [Fig.~\ref{incav}(a)] and $l=3.1$~$\mu$m [Fig.~\ref{incav}(b)]. The effective cavity lengths have been calculated from the free spectral range. Due to scattering losses induced by the ND the finesse is only $\mathcal{F}\approx 940$ corresponding to a linewidth (FWHM) of $\delta\nu\approx 46$~GHz. In Fig.~\ref{incav}(a) [Fig.~\ref{incav}(b)] the fundamental TEM$_{00}$ cavity mode is tuned to a spectral position close to the NV ZPL (PSB maximum). Saturation measurements of the emission into the fundamental mode yield $I_\infty = 770 (3700)$~counts/s and $P_\mathrm{sat} = 0.72(0.46)$~mW for $l=3.5(3.1)$~$\mu$m. Since the photons are emitted into the narrow cavity linewidth [$\delta \nu \approx 46 (90)$~GHz] we yield a high spectral photon density $> 17 (41)$~photons/(s~GHz) at the ZPL (PSB) wavelength which is significantly larger compared to the NV center emission into free space [$\approx 4.5$~photons/(s~GHz)]. This clearly proves cavity-enhanced emission beyond simple spectral filtering. This is even more evident if one additionally takes into account spatial filtering: the combined filtering effects would yield saturation count rates $\approx 2$~photons/s, $2-3$ orders of magnitude smaller than the observed rates~\cite{Supp}. The measured spectral photon densities are comparable to the brightest  emission of a single NV center reported so far~\cite{Schroeder11}: $2.4\times10^6$~counts/s into $\Delta\nu \approx 60$~THz corresponds to 40~photons/(s~GHz). As discussed above, the enhanced emission furthermore cannot be explained by a standard Purcell enhancement: for our parameters $g = 1.1$~GHz, $\gamma = 35$~MHz and $\gamma^\star = 15$~THz we expect $F \approx 10^{-2}$. This is confirmed by the experiment: The intensity correlation of the cavity emission (Fig.~\ref{g2meas}(b), recorded at $l=3.5$~$\mu$m) with $g^{(2)}(0)<0.5$ proves that indeed a single NV center is coupled to the cavity mode and by again deducing the parameters $a$, $\tau_1$, $\tau_2$ from $g^{(2)}$ functions at different pump powers [see Figs.~\ref{g2meas} (c)-(e)] we find that the internal dynamics are not modified by the cavity coupling.

\begin{figure}[htpb]
\centering\includegraphics[width=\columnwidth]{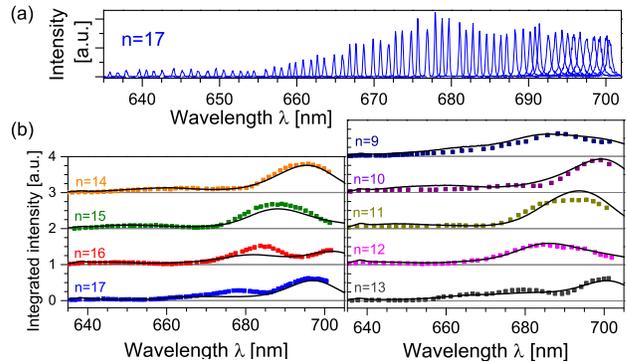}
\caption{(a) Intensity of TEM$_{00}$ mode for fixed longitudinal mode number $n=17$ at different effective cavity lengths $l = n \times \lambda/2$. (b) Dots: Area of each TEM$_{00}$ peak proportional to the photon count rates for $n=9-17$. Lines: Normalized count rates as obtained from the rate model.}
\label{tuning}
\end{figure}

Thus, our system must operate in a different coupling regime where the emitter's homogeneous broadening helps channeling emission into the cavity mode. We make use of the strong broadening by demonstrating tuning of the cavity over several longitudinal modes via continuously varying the mirror separation. At different cavity lengths $l = n \times \lambda/2$, where $n$ is a fixed longitudinal mode number, we record the intensity of the fundamental TEM$_{00}$ mode using a spectrometer. Figigure~\ref{tuning}(a) exemplarily shows a tuning spectrum for $n=17$, i.e. a sequence of successively recorded TEM$_{00}$ peaks. The area under each TEM$_{00}$ peak which is proportional to the photon rate emitted from the cavity is displayed as a function of wavelength for $n = 9-17$ in Fig.~\ref{tuning}(b). Note that due to residual reflectivity of the coatings also the excitation laser at $532$~nm builds up a standing wave inside the cavity leading to a strong modulation of mode intensities. As an important result we observe single photon emission into the cavity mode over a very broad spectral range, thereby realizing a continuously tunable, narrow band single photon source.

To explain our experimental results we expand a master equation model~\cite{Auffeves09} for solid-state emitter-cavity coupling. For the NV center we assume, in analogy to~\cite{Su08-2}, a simplified level scheme as depicted in Fig.~\ref{spectrum}(b), consisting of one excited state ($|e\rangle$) and $n$ vibronic ground states ($|g_0\rangle, \ldots, |g_n\rangle$), the energies of which are given by the position of the Lorentzians fitted to the spectrum of Fig.~\ref{spectrum}(a). The spontaneous decay rate on the ZPL ($|e\rangle \rightarrow |g_0\rangle$) is denoted by $\gamma_0$; on the PSB ($|e\rangle \rightarrow |g_i\rangle$) by $\gamma_i$. At room temperature there are two dominant line broadening mechanisms: first, all transitions are equally subject to phonon scattering which can be described by a pure dephasing rate $\gamma^\star$; second, the PSBs experience an additional radiative broadening $\gamma_{i,i-1}$  due to the short lifetime of the vibrational levels $|g_i\rangle$. From the fit to Fig.~\ref{spectrum}(a) we obtain $\gamma^\star = 15$~THz and $\gamma_{i,i-1}$ in the range $65-88$~THz. This level scheme represents a simplified description of the NV center PSB spectrum~\cite{Doherty13} but suffices to adequately account for the contributions of the individual transitions to the cavity coupling.

The coupling of the optical transitions to the cavity mode is described by a Jaynes-Cummings Hamiltonian:
\begin{eqnarray}
\hat{\mathcal{H}}^\mathrm{JC} & = & \sum_{i=0}^n\hbar\omega_i\hat{\sigma}_{i,i} + \hbar\omega_\mathrm{ZPL}\hat{\sigma}_{e,e} + \hbar\omega_c \hat{a}^\dagger\hat{a}  \nonumber\\
&& + i\hbar \sum_{i=0}^n g_i(\hat{a}^\dagger\hat{\sigma}_{i,e} - \hat{\sigma}^\dagger_{i,e}\hat{a})
\end{eqnarray}
where $\hbar\omega_i, i = 0,1,\ldots,n$ are the level energies (and $\omega_0 \equiv 0$), $\omega_\mathrm{ZPL}$ is the ZPL transition frequency, $\omega_c$ is the cavity resonance frequency, $g_i$ are the cavity coupling rates, $\hat{a}$ $(\hat{a}^\dagger)$ are cavity photon annihilation(creation) operators and $\hat{\sigma}_{n,m}$ are transition ($n\neq m$) or population ($n=m$) operators. Assuming Markov approximation the master equation for the coupled system is:
\begin{equation}
\dot{\rho} = \frac{i}{\hbar} \left[\rho, \hat{\mathcal{H}}^{JC}\right] + \mathcal{L}^\mathrm{cav}_\mathrm{damp} + \mathcal{L}^\mathrm{at}_\mathrm{damp} + \mathcal{L}^\mathrm{at}_\mathrm{GS Relax} + \mathcal{L}^\mathrm{at}_\mathrm{deph}
\end{equation}
taking into account cavity loss with a rate $\kappa$ ($\mathcal{L}^\mathrm{cav}_\mathrm{damp}$), loss of polarization ($\mathcal{L}^\mathrm{at}_\mathrm{damp}$) and broadening effects as discussed above ($\mathcal{L}^\mathrm{at}_\mathrm{GS Relax} + \mathcal{L}^\mathrm{at}_\mathrm{deph}$). The master equation can be used to calculate the full time evolution of populations and coherences. At room temperature, however, line broadening dominates all other rates and coherences can be adiabatically eliminated leading to a rate equation system~\cite{Supp}. In analogy to~\cite{Auffeves09} the rate equations' dynamics are governed by effective coupling rates between the emitter's transition and the cavity mode:
\begin{subequations}
\begin{align}
R_0 &= \frac{4g_0^2}{\kappa+\gamma +\gamma^\star}\frac{1}{1+\left(\frac{2\delta_0}{\kappa+\gamma+\gamma^\star}\right)^2}\\
R_i &= \frac{4g_i^2}{\kappa+\gamma+\gamma_{i,i-1}+\gamma^\star}\frac{1}{1+\left(\frac{2\delta_1}{\kappa+\gamma+\gamma_{i,i-1}+\gamma^\star}\right)^2}
\end{align}
\end{subequations}
where $i=1,\ldots,n$, $\gamma = \gamma_0+\gamma_1+\ldots+\gamma_n$ and $\delta_i = \omega_i - \omega_c$. For fast cavity decay $\kappa >> \gamma,R_i$ we determine the emission efficiency $P_i$ of each transition as ratio of the emission coupled into the cavity to the total NV emission: $P_i = R_i/(\gamma + \sum_{j=0}^n{R_j}), \quad i = 0,\ldots,n$. If the cavity is tuned to any arbitrary wavelength $\lambda$ within the NV emission spectrum, the total emission into the cavity mode is the sum of contributions from ZPL and all PSB: $P_\mathrm{tot}(\lambda) = \sum_i P_i(\lambda) = \sum_iR_i(\lambda)/\big(\gamma+\sum_iR_i(\lambda)\big)$.

\begin{figure}[htbp]
\centering\includegraphics[width=\columnwidth]{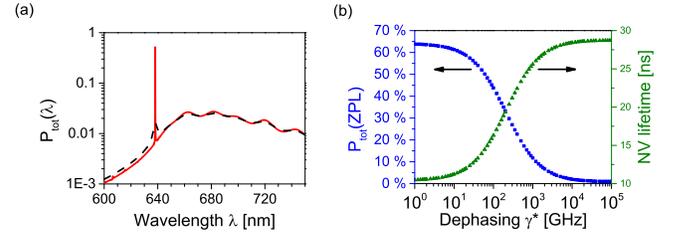}
\caption{(a) Simulated emission efficiency $P_\mathrm{tot}(\lambda)$ as a function of wavelength for two dephasing rates, $\gamma^\star = 15$~THz (dashed line) and $\gamma^\star = 30$~GHz (solid line), corresponding to room and cryogenic temperatures, respectively. (b) NV spontaneous emission lifetime and emission efficiency $P_\mathrm{tot}(\mathrm{ZPL})$ at the ZPL wavelength as a function of the pure dephasing rate $\gamma^\star$. For both (a) and (b) $\kappa =77$~GHz ($\mathcal{F}=3500$) and $l = 9/2\times \lambda$.}
\label{cold-theory}
\end{figure}

All the parameters of our model can be determined experimentally~\cite{Supp}. The model can be used to predict absolute emitted photon numbers for a cavity mode tuned to a fixed wavelength [as in Figs.~\ref{incav}(a),(b)]. With efficiencies $P_\mathrm{tot} = 1.37$\%($3.35$\%) derived from the model for effective cavity lengths $l= 3.5(3.1)$~$\mu$m we predict count rates of $530(3200)$~s$^{-1}$ which favorably compares to the saturation measurement results $I_\infty = 770(3700)$~s$^{-1}$. In the prediction we included the wavelength-dependent output coupling into the fiber~\cite{Supp}.
Similarly, the rate equation model qualitatively describes the tuning spectra of Fig.~\ref{tuning}(b). We first calculate the efficiencies $P_\mathrm{tot}(\lambda)$ for emission at different wavelengths for the given cavity parameters. Figure~\ref{cold-theory}(a) shows that at room temperature $P_\mathrm{tot}(\lambda)$ only varies slightly over the NV emission spectrum. Secondly, we have to take into account the modulation due to the excitation light standing wave and the wavelength dependence of finesse and outcoupling from the cavity to the fiber mode. From a simultaneous fit to all data of Fig.~\ref{tuning}(b) we determine a loss rate of $0.54$\% due to scattering from the ND~\cite{Supp}.

The full master equation model furthermore can be used to predict the low-temperature coupling dynamics of the NV center and fiber cavity. When - for decreasing temperatures - the total ZPL linewidth ($\gamma_0 + \gamma^\star$) approaches the cavity linewidth a transition from dephasing-assisted emission to Purcell-enhanced emission occurs. Figure~\ref{cold-theory}(b) shows this transition as a variation of emission efficiency $P_\mathrm{tot}(\lambda_\mathrm{ZPL})$ at the ZPL wavelength and the NV center spontaneous emission lifetime as a function of the pure dephasing rate. For large $\gamma^\star$ we recover the room-temperature situation where the emission lifetime is only slightly modified and the emission efficiencies are in the range of a few percent. For $\gamma^\star < 10$~GHz, the lifetime drastically reduces by a factor of $\approx 3$ and up to $65$\% of the NV emission are channeled into the cavity mode. Figure~\ref{cold-theory}(a) shows that the emission enhancement is effective only at the ZPL position whereas the PSB emission is almost unaltered.

In summary, by employing a fiber-based microcavity we have explored phonon-assisted cavity feeding, a coupling scheme previously unknown for the NV center in diamond. At any given wavelength within the NV emission spectrum the ZPL and all phonon sidebands contribute to the cavity emission enabling the demonstration of a widely tunable, narrow band single photon source. The single photon emission efficiency and tuning spectra are well described by a theoretical model taking into account the NV center phonon broadening. For low temperatures our model predicts a transition into the Purcell regime where up to $65$\% of the NV emission would be channeled into the cavity mode for our given experimental parameters. Currently, the cavity finesse is limited by scattering losses from the ND. By employing smaller NDs an increase in finesse by a factor $> 10$ should be possible, enabling schemes for cavity-enhanced spin measurements~\cite{Young09} or creation of entangled states~\cite{Young12}. The large photon collection efficiency of the coupled NV-cavity system could also boost protocols for heralded entanglement of remote qubits~\cite{Bernien12}. On the other hand, the cavity mode volume can be reduced significantly to values as small as $\approx 0.6$~$\mu$m$^3$ by fabricating mirrors with radii of curvature down to $5$~$\mu$m, accessible by focused ion beam milling. Together with a finesse of $\mathcal{F}=10\:000$ this would enable a widely tunable narrow ($10$~GHz) room-temperature single photon source with count rates $> 10^5$~counts/s.

\begin{acknowledgments}
The authors acknowledge helpful discussions with D. Hunger, M. Bienert and J. Neergaard-Nielsen and thank B. Sauer for help in the early stage of the experiment. The project was financially supported by the Bundesministerium f\"ur Bildung und Forschung within the projects EPHQUAM (Contract No. 01BL0903) and QuOReP (Contract NO. 01BQ1011).
\end{acknowledgments}




\begin{thebibliography}{27}%
\makeatletter
\providecommand \@ifxundefined [1]{%
 \@ifx{#1\undefined}
}%
\providecommand \@ifnum [1]{%
 \ifnum #1\expandafter \@firstoftwo
 \else \expandafter \@secondoftwo
 \fi
}%
\providecommand \@ifx [1]{%
 \ifx #1\expandafter \@firstoftwo
 \else \expandafter \@secondoftwo
 \fi
}%
\providecommand \natexlab [1]{#1}%
\providecommand \enquote  [1]{``#1''}%
\providecommand \bibnamefont  [1]{#1}%
\providecommand \bibfnamefont [1]{#1}%
\providecommand \citenamefont [1]{#1}%
\providecommand \href@noop [0]{\@secondoftwo}%
\providecommand \href [0]{\begingroup \@sanitize@url \@href}%
\providecommand \@href[1]{\@@startlink{#1}\@@href}%
\providecommand \@@href[1]{\endgroup#1\@@endlink}%
\providecommand \@sanitize@url [0]{\catcode `\\12\catcode `\$12\catcode
  `\&12\catcode `\#12\catcode `\^12\catcode `\_12\catcode `\%12\relax}%
\providecommand \@@startlink[1]{}%
\providecommand \@@endlink[0]{}%
\providecommand \url  [0]{\begingroup\@sanitize@url \@url }%
\providecommand \@url [1]{\endgroup\@href {#1}{\urlprefix }}%
\providecommand \urlprefix  [0]{URL }%
\providecommand \Eprint [0]{\href }%
\providecommand \doibase [0]{http://dx.doi.org/}%
\providecommand \selectlanguage [0]{\@gobble}%
\providecommand \bibinfo  [0]{\@secondoftwo}%
\providecommand \bibfield  [0]{\@secondoftwo}%
\providecommand \translation [1]{[#1]}%
\providecommand \BibitemOpen [0]{}%
\providecommand \bibitemStop [0]{}%
\providecommand \bibitemNoStop [0]{.\EOS\space}%
\providecommand \EOS [0]{\spacefactor3000\relax}%
\providecommand \BibitemShut  [1]{\csname bibitem#1\endcsname}%
\let\auto@bib@innerbib\@empty
\bibitem [{\citenamefont {Kurtsiefer}\ \emph {et~al.}(2000)\citenamefont
  {Kurtsiefer}, \citenamefont {Mayer}, \citenamefont {Zarda},\ and\
  \citenamefont {Weinfurter}}]{Kurtsiefer00}%
  \BibitemOpen
  \bibfield  {author} {\bibinfo {author} {\bibfnamefont {C.}~\bibnamefont
  {Kurtsiefer}}, \bibinfo {author} {\bibfnamefont {S.}~\bibnamefont {Mayer}},
  \bibinfo {author} {\bibfnamefont {P.}~\bibnamefont {Zarda}}, \ and\ \bibinfo
  {author} {\bibfnamefont {H.}~\bibnamefont {Weinfurter}},\ }\href@noop {}
  {\bibfield  {journal} {\bibinfo  {journal} {Phys. Rev. Lett.}\ }\textbf
  {\bibinfo {volume} {85}},\ \bibinfo {pages} {290} (\bibinfo {year}
  {2000})}\BibitemShut {NoStop}%
\bibitem [{\citenamefont {Neu}\ \emph {et~al.}(2011)\citenamefont {Neu},
  \citenamefont {Steinmetz}, \citenamefont {Riedrich-M\"oller}, \citenamefont
  {Gsell}, \citenamefont {Fischer}, \citenamefont {Schreck},\ and\
  \citenamefont {Becher}}]{Neu11}%
  \BibitemOpen
  \bibfield  {author} {\bibinfo {author} {\bibfnamefont {E.}~\bibnamefont
  {Neu}}, \bibinfo {author} {\bibfnamefont {D.}~\bibnamefont {Steinmetz}},
  \bibinfo {author} {\bibfnamefont {J.}~\bibnamefont {Riedrich-M\"oller}},
  \bibinfo {author} {\bibfnamefont {S.}~\bibnamefont {Gsell}}, \bibinfo
  {author} {\bibfnamefont {M.}~\bibnamefont {Fischer}}, \bibinfo {author}
  {\bibfnamefont {M.}~\bibnamefont {Schreck}}, \ and\ \bibinfo {author}
  {\bibfnamefont {C.}~\bibnamefont {Becher}},\ }\href {\doibase
  10.1088/1367-2630/13/2/025012} {\bibfield  {journal} {\bibinfo  {journal}
  {New J. Phys.}\ }\textbf {\bibinfo {volume} {13}},\ \bibinfo {pages} {025012}
  (\bibinfo {year} {2011})}\BibitemShut {NoStop}%
\bibitem [{\citenamefont {Prawer}\ and\ \citenamefont
  {Greentree}(2008)}]{Prawer08}%
  \BibitemOpen
  \bibfield  {author} {\bibinfo {author} {\bibfnamefont {S.}~\bibnamefont
  {Prawer}}\ and\ \bibinfo {author} {\bibfnamefont {A.~D.}\ \bibnamefont
  {Greentree}},\ }\href {\doibase 10.1126/science.1158340} {\bibfield
  {journal} {\bibinfo  {journal} {Science}\ }\textbf {\bibinfo {volume}
  {320}},\ \bibinfo {pages} {1601} (\bibinfo {year} {2008})}\BibitemShut
  {NoStop}%
\bibitem [{\citenamefont {Aharonovich}\ \emph {et~al.}(2011)\citenamefont
  {Aharonovich}, \citenamefont {Greentree},\ and\ \citenamefont
  {Prawer}}]{Aharonovich11}%
  \BibitemOpen
  \bibfield  {author} {\bibinfo {author} {\bibfnamefont {I.}~\bibnamefont
  {Aharonovich}}, \bibinfo {author} {\bibfnamefont {A.~D.}\ \bibnamefont
  {Greentree}}, \ and\ \bibinfo {author} {\bibfnamefont {S.}~\bibnamefont
  {Prawer}},\ }\href {\doibase 10.1038/NPHOTON.2011.54} {\bibfield  {journal}
  {\bibinfo  {journal} {Nat. Photonics}\ }\textbf {\bibinfo {volume} {5}},\
  \bibinfo {pages} {397} (\bibinfo {year} {2011})}\BibitemShut {NoStop}%
\bibitem [{\citenamefont {Benjamin}\ \emph {et~al.}(2009)\citenamefont
  {Benjamin}, \citenamefont {Lovett},\ and\ \citenamefont
  {Smith}}]{Benjamin09}%
  \BibitemOpen
  \bibfield  {author} {\bibinfo {author} {\bibfnamefont {S.~C.}\ \bibnamefont
  {Benjamin}}, \bibinfo {author} {\bibfnamefont {B.~W.}\ \bibnamefont
  {Lovett}}, \ and\ \bibinfo {author} {\bibfnamefont {J.~M.}\ \bibnamefont
  {Smith}},\ }\href {\doibase 10.1002/lpor.200810051} {\bibfield  {journal}
  {\bibinfo  {journal} {Laser Photonics Rev}\ }\textbf {\bibinfo {volume}
  {3}},\ \bibinfo {pages} {556} (\bibinfo {year} {2009})}\BibitemShut {NoStop}%
\bibitem [{\citenamefont {Acosta}\ \emph {et~al.}(2010)\citenamefont {Acosta},
  \citenamefont {Bauch}, \citenamefont {Jarmola}, \citenamefont {Zipp},
  \citenamefont {Ledbetter},\ and\ \citenamefont {Budker}}]{Acosta10}%
  \BibitemOpen
  \bibfield  {author} {\bibinfo {author} {\bibfnamefont {V.~M.}\ \bibnamefont
  {Acosta}}, \bibinfo {author} {\bibfnamefont {E.}~\bibnamefont {Bauch}},
  \bibinfo {author} {\bibfnamefont {A.}~\bibnamefont {Jarmola}}, \bibinfo
  {author} {\bibfnamefont {L.~J.}\ \bibnamefont {Zipp}}, \bibinfo {author}
  {\bibfnamefont {M.~P.}\ \bibnamefont {Ledbetter}}, \ and\ \bibinfo {author}
  {\bibfnamefont {D.}~\bibnamefont {Budker}},\ }\href {\doibase
  10.1063/1.3507884} {\bibfield  {journal} {\bibinfo  {journal} {Appl. Phys.
  Lett.}\ }\textbf {\bibinfo {volume} {97}},\ \bibinfo {pages} {174104}
  (\bibinfo {year} {2010})}\BibitemShut {NoStop}%
\bibitem [{\citenamefont {Su}\ \emph {et~al.}(2008)\citenamefont {Su},
  \citenamefont {Greentree},\ and\ \citenamefont {Hollenberg}}]{Su08-2}%
  \BibitemOpen
  \bibfield  {author} {\bibinfo {author} {\bibfnamefont {C.-H.}\ \bibnamefont
  {Su}}, \bibinfo {author} {\bibfnamefont {A.~D.}\ \bibnamefont {Greentree}}, \
  and\ \bibinfo {author} {\bibfnamefont {L.~C.~L.}\ \bibnamefont
  {Hollenberg}},\ }\href@noop {} {\bibfield  {journal} {\bibinfo  {journal}
  {Opt. Express}\ }\textbf {\bibinfo {volume} {16}},\ \bibinfo {pages} {6240}
  (\bibinfo {year} {2008})}\BibitemShut {NoStop}%
\bibitem [{\citenamefont {Young}\ \emph {et~al.}(2009)\citenamefont {Young},
  \citenamefont {Hu}, \citenamefont {Marseglia}, \citenamefont {Harrison},
  \citenamefont {O'Brien},\ and\ \citenamefont {Rarity}}]{Young09}%
  \BibitemOpen
  \bibfield  {author} {\bibinfo {author} {\bibfnamefont {A.}~\bibnamefont
  {Young}}, \bibinfo {author} {\bibfnamefont {C.~Y.}\ \bibnamefont {Hu}},
  \bibinfo {author} {\bibfnamefont {L.}~\bibnamefont {Marseglia}}, \bibinfo
  {author} {\bibfnamefont {J.~P.}\ \bibnamefont {Harrison}}, \bibinfo {author}
  {\bibfnamefont {J.~L.}\ \bibnamefont {O'Brien}}, \ and\ \bibinfo {author}
  {\bibfnamefont {J.~G.}\ \bibnamefont {Rarity}},\ }\href {\doibase
  10.1088/1367-2630/11/1/013007} {\bibfield  {journal} {\bibinfo  {journal}
  {New J. Phys.}\ }\textbf {\bibinfo {volume} {11}},\ \bibinfo {pages} {013007}
  (\bibinfo {year} {2009})}\BibitemShut {NoStop}%
\bibitem [{\citenamefont {Riedrich-M\"oller}\ \emph {et~al.}(2012)\citenamefont
  {Riedrich-M\"oller}, \citenamefont {Kipfstuhl}, \citenamefont {Hepp},
  \citenamefont {Neu}, \citenamefont {Pauly}, \citenamefont {M\"ucklich},
  \citenamefont {Baur}, \citenamefont {Wandt}, \citenamefont {Wolff},
  \citenamefont {Fischer}, \citenamefont {Gsell}, \citenamefont {Schreck},\
  and\ \citenamefont {Becher}}]{RiedrichMoeller12}%
  \BibitemOpen
  \bibfield  {author} {\bibinfo {author} {\bibfnamefont {J.}~\bibnamefont
  {Riedrich-M\"oller}}, \bibinfo {author} {\bibfnamefont {L.}~\bibnamefont
  {Kipfstuhl}}, \bibinfo {author} {\bibfnamefont {C.}~\bibnamefont {Hepp}},
  \bibinfo {author} {\bibfnamefont {E.}~\bibnamefont {Neu}}, \bibinfo {author}
  {\bibfnamefont {C.}~\bibnamefont {Pauly}}, \bibinfo {author} {\bibfnamefont
  {F.}~\bibnamefont {M\"ucklich}}, \bibinfo {author} {\bibfnamefont
  {A.}~\bibnamefont {Baur}}, \bibinfo {author} {\bibfnamefont {M.}~\bibnamefont
  {Wandt}}, \bibinfo {author} {\bibfnamefont {S.}~\bibnamefont {Wolff}},
  \bibinfo {author} {\bibfnamefont {M.}~\bibnamefont {Fischer}}, \bibinfo
  {author} {\bibfnamefont {S.}~\bibnamefont {Gsell}}, \bibinfo {author}
  {\bibfnamefont {M.}~\bibnamefont {Schreck}}, \ and\ \bibinfo {author}
  {\bibfnamefont {C.}~\bibnamefont {Becher}},\ }\href {\doibase
  10.1038/NNANO.2011.190} {\bibfield  {journal} {\bibinfo  {journal} {Nature
  Nanotech.}\ }\textbf {\bibinfo {volume} {7}},\ \bibinfo {pages} {69}
  (\bibinfo {year} {2012})}\BibitemShut {NoStop}%
\bibitem [{\citenamefont {Faraon}\ \emph {et~al.}(2012)\citenamefont {Faraon},
  \citenamefont {Santori}, \citenamefont {Huang}, \citenamefont {Acosta},\ and\
  \citenamefont {Beausoleil}}]{Faraon12}%
  \BibitemOpen
  \bibfield  {author} {\bibinfo {author} {\bibfnamefont {A.}~\bibnamefont
  {Faraon}}, \bibinfo {author} {\bibfnamefont {C.}~\bibnamefont {Santori}},
  \bibinfo {author} {\bibfnamefont {Z.}~\bibnamefont {Huang}}, \bibinfo
  {author} {\bibfnamefont {V.~M.}\ \bibnamefont {Acosta}}, \ and\ \bibinfo
  {author} {\bibfnamefont {R.~G.}\ \bibnamefont {Beausoleil}},\ }\href
  {\doibase 10.1103/PhysRevLett.109.033604} {\bibfield  {journal} {\bibinfo
  {journal} {Phys. Rev. Lett.}\ }\textbf {\bibinfo {volume} {109}},\ \bibinfo
  {pages} {033604} (\bibinfo {year} {2012})}\BibitemShut {NoStop}%
\bibitem [{\citenamefont {Colombe}\ \emph {et~al.}(2007)\citenamefont
  {Colombe}, \citenamefont {Steinmetz}, \citenamefont {Dubois}, \citenamefont
  {Linke}, \citenamefont {Hunger},\ and\ \citenamefont {Reichel}}]{Colombe07}%
  \BibitemOpen
  \bibfield  {author} {\bibinfo {author} {\bibfnamefont {Y.}~\bibnamefont
  {Colombe}}, \bibinfo {author} {\bibfnamefont {T.}~\bibnamefont {Steinmetz}},
  \bibinfo {author} {\bibfnamefont {G.}~\bibnamefont {Dubois}}, \bibinfo
  {author} {\bibfnamefont {F.}~\bibnamefont {Linke}}, \bibinfo {author}
  {\bibfnamefont {D.}~\bibnamefont {Hunger}}, \ and\ \bibinfo {author}
  {\bibfnamefont {J.}~\bibnamefont {Reichel}},\ }\href {\doibase
  10.1038/nature06331} {\bibfield  {journal} {\bibinfo  {journal} {Nature}\
  }\textbf {\bibinfo {volume} {450}},\ \bibinfo {pages} {272} (\bibinfo {year}
  {2007})}\BibitemShut {NoStop}%
\bibitem [{\citenamefont {Hunger}\ \emph {et~al.}(2010)\citenamefont {Hunger},
  \citenamefont {Steinmetz}, \citenamefont {Colombe}, \citenamefont {Deutsch},
  \citenamefont {H\"ansch},\ and\ \citenamefont {Reichel}}]{Hunger10}%
  \BibitemOpen
  \bibfield  {author} {\bibinfo {author} {\bibfnamefont {D.}~\bibnamefont
  {Hunger}}, \bibinfo {author} {\bibfnamefont {T.}~\bibnamefont {Steinmetz}},
  \bibinfo {author} {\bibfnamefont {Y.}~\bibnamefont {Colombe}}, \bibinfo
  {author} {\bibfnamefont {C.}~\bibnamefont {Deutsch}}, \bibinfo {author}
  {\bibfnamefont {T.~W.}\ \bibnamefont {H\"ansch}}, \ and\ \bibinfo {author}
  {\bibfnamefont {J.}~\bibnamefont {Reichel}},\ }\href {\doibase
  10.1088/1367-2630/12/6/065038} {\bibfield  {journal} {\bibinfo  {journal}
  {New J. Phys.}\ }\textbf {\bibinfo {volume} {12}},\ \bibinfo {pages} {065038}
  (\bibinfo {year} {2010})}\BibitemShut {NoStop}%
\bibitem [{\citenamefont {Muller}\ \emph {et~al.}(2009)\citenamefont {Muller},
  \citenamefont {Flagg}, \citenamefont {Metcalfe}, \citenamefont {Lawall},\
  and\ \citenamefont {Solomon}}]{Muller09}%
  \BibitemOpen
  \bibfield  {author} {\bibinfo {author} {\bibfnamefont {A.}~\bibnamefont
  {Muller}}, \bibinfo {author} {\bibfnamefont {E.~B.}\ \bibnamefont {Flagg}},
  \bibinfo {author} {\bibfnamefont {M.}~\bibnamefont {Metcalfe}}, \bibinfo
  {author} {\bibfnamefont {J.}~\bibnamefont {Lawall}}, \ and\ \bibinfo {author}
  {\bibfnamefont {G.~S.}\ \bibnamefont {Solomon}},\ }\href {\doibase
  10.1063/1.3245311} {\bibfield  {journal} {\bibinfo  {journal} {Appl. Phys.
  Lett.}\ }\textbf {\bibinfo {volume} {95}},\ \bibinfo {pages} {173101}
  (\bibinfo {year} {2009})}\BibitemShut {NoStop}%
\bibitem [{\citenamefont {Miguel-Sanch\'ez}\ \emph {et~al.}(2013)\citenamefont
  {Miguel-Sanch\'ez}, \citenamefont {Reinhard}, \citenamefont {Togan},
  \citenamefont {Volz}, \citenamefont {Imamo\u{g}lu}, \citenamefont {Besga},
  \citenamefont {Reichel},\ and\ \citenamefont {Est\`eve}}]{Sanchez12}%
  \BibitemOpen
  \bibfield  {author} {\bibinfo {author} {\bibfnamefont {J.}~\bibnamefont
  {Miguel-Sanch\'ez}}, \bibinfo {author} {\bibfnamefont {A.}~\bibnamefont
  {Reinhard}}, \bibinfo {author} {\bibfnamefont {E.}~\bibnamefont {Togan}},
  \bibinfo {author} {\bibfnamefont {T.}~\bibnamefont {Volz}}, \bibinfo {author}
  {\bibfnamefont {A.}~\bibnamefont {Imamo\u{g}lu}}, \bibinfo {author}
  {\bibfnamefont {B.}~\bibnamefont {Besga}}, \bibinfo {author} {\bibfnamefont
  {J.}~\bibnamefont {Reichel}}, \ and\ \bibinfo {author} {\bibfnamefont
  {J.}~\bibnamefont {Est\`eve}},\ }\href
  {http://stacks.iop.org/1367-2630/15/i=4/a=045002} {\bibfield  {journal}
  {\bibinfo  {journal} {New J. Phys.}\ }\textbf {\bibinfo {volume} {15}},\
  \bibinfo {pages} {045002} (\bibinfo {year} {2013})}\BibitemShut {NoStop}%
\bibitem [{\citenamefont {Kaupp}\ \emph {et~al.}(2013)\citenamefont {Kaupp},
  \citenamefont {Deutsch}, \citenamefont {Chang}, \citenamefont {Reichel},
  \citenamefont {H\"ansch},\ and\ \citenamefont {Hunger}}]{Kaupp13}%
  \BibitemOpen
  \bibfield  {author} {\bibinfo {author} {\bibfnamefont {H.}~\bibnamefont
  {Kaupp}}, \bibinfo {author} {\bibfnamefont {C.}~\bibnamefont {Deutsch}},
  \bibinfo {author} {\bibfnamefont {H.-C.}\ \bibnamefont {Chang}}, \bibinfo
  {author} {\bibfnamefont {J.}~\bibnamefont {Reichel}}, \bibinfo {author}
  {\bibfnamefont {T.~W.}\ \bibnamefont {H\"ansch}}, \ and\ \bibinfo {author}
  {\bibfnamefont {D.}~\bibnamefont {Hunger}},\ }\href@noop {} {\bibfield
  {journal} {\bibinfo  {journal} {arXiv:1304.0948}\ } (\bibinfo {year}
  {2013})}\BibitemShut {NoStop}%
\bibitem [{\citenamefont {Auff\`eves}\ \emph {et~al.}(2009)\citenamefont
  {Auff\`eves}, \citenamefont {G\'erard},\ and\ \citenamefont
  {Poizat}}]{Auffeves09}%
  \BibitemOpen
  \bibfield  {author} {\bibinfo {author} {\bibfnamefont {A.}~\bibnamefont
  {Auff\`eves}}, \bibinfo {author} {\bibfnamefont {J.-M.}\ \bibnamefont
  {G\'erard}}, \ and\ \bibinfo {author} {\bibfnamefont {J.-P.}\ \bibnamefont
  {Poizat}},\ }\href {\doibase 10.1103/PhysRevA.79.053838} {\bibfield
  {journal} {\bibinfo  {journal} {Phys. Rev. A}\ }\textbf {\bibinfo {volume}
  {79}},\ \bibinfo {pages} {053838} (\bibinfo {year} {2009})}\BibitemShut
  {NoStop}%
\bibitem [{\citenamefont {Auff\`eves}\ \emph {et~al.}(2010)\citenamefont
  {Auff\`eves}, \citenamefont {Gerace}, \citenamefont {G\'erard}, \citenamefont
  {Santos}, \citenamefont {Andreani},\ and\ \citenamefont
  {Poizat}}]{Auffeves10}%
  \BibitemOpen
  \bibfield  {author} {\bibinfo {author} {\bibfnamefont {A.}~\bibnamefont
  {Auff\`eves}}, \bibinfo {author} {\bibfnamefont {D.}~\bibnamefont {Gerace}},
  \bibinfo {author} {\bibfnamefont {J.-M.}\ \bibnamefont {G\'erard}}, \bibinfo
  {author} {\bibfnamefont {M.~F.}\ \bibnamefont {Santos}}, \bibinfo {author}
  {\bibfnamefont {L.~C.}\ \bibnamefont {Andreani}}, \ and\ \bibinfo {author}
  {\bibfnamefont {J.-P.}\ \bibnamefont {Poizat}},\ }\href {\doibase
  10.1103/PhysRevB.81.245419} {\bibfield  {journal} {\bibinfo  {journal} {Phys.
  Rev. B}\ }\textbf {\bibinfo {volume} {81}},\ \bibinfo {pages} {245419}
  (\bibinfo {year} {2010})}\BibitemShut {NoStop}%
\bibitem [{\citenamefont {Wilson-Rae}\ and\ \citenamefont
  {Imamo\u{g}lu}(2002)}]{Wilson-Rae02}%
  \BibitemOpen
  \bibfield  {author} {\bibinfo {author} {\bibfnamefont {I.}~\bibnamefont
  {Wilson-Rae}}\ and\ \bibinfo {author} {\bibfnamefont {A.}~\bibnamefont
  {Imamo\u{g}lu}},\ }\href {\doibase 10.1103/PhysRevB.65.235311} {\bibfield
  {journal} {\bibinfo  {journal} {Phys. Rev. B}\ }\textbf {\bibinfo {volume}
  {65}},\ \bibinfo {pages} {235311} (\bibinfo {year} {2002})}\BibitemShut
  {NoStop}%
\bibitem [{\citenamefont {Hohenester}\ \emph {et~al.}(2009)\citenamefont
  {Hohenester}, \citenamefont {Laucht}, \citenamefont {Kaniber}, \citenamefont
  {Hauke}, \citenamefont {Neumann}, \citenamefont {Mohtashami}, \citenamefont
  {Seliger}, \citenamefont {Bichler},\ and\ \citenamefont
  {Finley}}]{Hohenester09}%
  \BibitemOpen
  \bibfield  {author} {\bibinfo {author} {\bibfnamefont {U.}~\bibnamefont
  {Hohenester}}, \bibinfo {author} {\bibfnamefont {A.}~\bibnamefont {Laucht}},
  \bibinfo {author} {\bibfnamefont {M.}~\bibnamefont {Kaniber}}, \bibinfo
  {author} {\bibfnamefont {N.}~\bibnamefont {Hauke}}, \bibinfo {author}
  {\bibfnamefont {A.}~\bibnamefont {Neumann}}, \bibinfo {author} {\bibfnamefont
  {A.}~\bibnamefont {Mohtashami}}, \bibinfo {author} {\bibfnamefont
  {M.}~\bibnamefont {Seliger}}, \bibinfo {author} {\bibfnamefont
  {M.}~\bibnamefont {Bichler}}, \ and\ \bibinfo {author} {\bibfnamefont
  {J.~J.}\ \bibnamefont {Finley}},\ }\href {\doibase
  10.1103/PhysRevB.80.201311} {\bibfield  {journal} {\bibinfo  {journal} {Phys.
  Rev. B}\ }\textbf {\bibinfo {volume} {80}},\ \bibinfo {pages} {201311}
  (\bibinfo {year} {2009})}\BibitemShut {NoStop}%
\bibitem [{\citenamefont {Ates}\ \emph {et~al.}(2009)\citenamefont {Ates},
  \citenamefont {Ulrich}, \citenamefont {Ulhaq}, \citenamefont {Reitzenstein},
  \citenamefont {L\"offler}, \citenamefont {H\"ofling}, \citenamefont
  {Forchel},\ and\ \citenamefont {Michler}}]{Ates09}%
  \BibitemOpen
  \bibfield  {author} {\bibinfo {author} {\bibfnamefont {S.}~\bibnamefont
  {Ates}}, \bibinfo {author} {\bibfnamefont {S.~M.}\ \bibnamefont {Ulrich}},
  \bibinfo {author} {\bibfnamefont {A.}~\bibnamefont {Ulhaq}}, \bibinfo
  {author} {\bibfnamefont {S.}~\bibnamefont {Reitzenstein}}, \bibinfo {author}
  {\bibfnamefont {A.}~\bibnamefont {L\"offler}}, \bibinfo {author}
  {\bibfnamefont {S.}~\bibnamefont {H\"ofling}}, \bibinfo {author}
  {\bibfnamefont {A.}~\bibnamefont {Forchel}}, \ and\ \bibinfo {author}
  {\bibfnamefont {P.}~\bibnamefont {Michler}},\ }\href {\doibase
  10.1038/nphoton.2009.215} {\bibfield  {journal} {\bibinfo  {journal} {Nat.
  Photonics}\ }\textbf {\bibinfo {volume} {3}},\ \bibinfo {pages} {724}
  (\bibinfo {year} {2009})}\BibitemShut {NoStop}%
\bibitem [{\citenamefont {Hennessy}\ \emph {et~al.}(2007)\citenamefont
  {Hennessy}, \citenamefont {Badolato}, \citenamefont {Winger}, \citenamefont
  {Gerace}, \citenamefont {Atat\"ure}, \citenamefont {Gulde}, \citenamefont
  {Faelt}, \citenamefont {Hu},\ and\ \citenamefont
  {Imamo\u{g}lu}}]{Hennessy07}%
  \BibitemOpen
  \bibfield  {author} {\bibinfo {author} {\bibfnamefont {K.}~\bibnamefont
  {Hennessy}}, \bibinfo {author} {\bibfnamefont {A.}~\bibnamefont {Badolato}},
  \bibinfo {author} {\bibfnamefont {M.}~\bibnamefont {Winger}}, \bibinfo
  {author} {\bibfnamefont {D.}~\bibnamefont {Gerace}}, \bibinfo {author}
  {\bibfnamefont {M.}~\bibnamefont {Atat\"ure}}, \bibinfo {author}
  {\bibfnamefont {S.}~\bibnamefont {Gulde}}, \bibinfo {author} {\bibfnamefont
  {S.}~\bibnamefont {Faelt}}, \bibinfo {author} {\bibfnamefont {E.~L.}\
  \bibnamefont {Hu}}, \ and\ \bibinfo {author} {\bibfnamefont {A.}~\bibnamefont
  {Imamo\u{g}lu}},\ }\href {\doibase 10.1038/nature05586} {\bibfield  {journal}
  {\bibinfo  {journal} {Nature}\ }\textbf {\bibinfo {volume} {445}},\ \bibinfo
  {pages} {896} (\bibinfo {year} {2007})}\BibitemShut {NoStop}%
\bibitem [{\citenamefont {Press}\ \emph {et~al.}(2007)\citenamefont {Press},
  \citenamefont {G\"otzinger}, \citenamefont {Reitzenstein}, \citenamefont
  {Hofmann}, \citenamefont {L\"offler}, \citenamefont {Kamp}, \citenamefont
  {Forchel},\ and\ \citenamefont {Yamamoto}}]{Press07}%
  \BibitemOpen
  \bibfield  {author} {\bibinfo {author} {\bibfnamefont {D.}~\bibnamefont
  {Press}}, \bibinfo {author} {\bibfnamefont {S.}~\bibnamefont {G\"otzinger}},
  \bibinfo {author} {\bibfnamefont {S.}~\bibnamefont {Reitzenstein}}, \bibinfo
  {author} {\bibfnamefont {C.}~\bibnamefont {Hofmann}}, \bibinfo {author}
  {\bibfnamefont {A.}~\bibnamefont {L\"offler}}, \bibinfo {author}
  {\bibfnamefont {M.}~\bibnamefont {Kamp}}, \bibinfo {author} {\bibfnamefont
  {A.}~\bibnamefont {Forchel}}, \ and\ \bibinfo {author} {\bibfnamefont
  {Y.}~\bibnamefont {Yamamoto}},\ }\href {\doibase
  10.1103/PhysRevLett.98.117402} {\bibfield  {journal} {\bibinfo  {journal}
  {Phys. Rev. Lett.}\ }\textbf {\bibinfo {volume} {98}},\ \bibinfo {pages}
  {117402} (\bibinfo {year} {2007})}\BibitemShut {NoStop}%
\bibitem [{Sup()}]{Supp}%
  \BibitemOpen
  \href@noop {} {}\bibinfo {note} {See Supplemental Material at [URL will be
  inserted by publisher] for details on experimental methods, data analysis,
  derivation of the theoretical model and all experimental
  parameters.}\BibitemShut {Stop}%
\bibitem [{\citenamefont {Schr\"oder}\ \emph {et~al.}(2011)\citenamefont
  {Schr\"oder}, \citenamefont {G\"adeke}, \citenamefont {Banholzer},\ and\
  \citenamefont {Benson}}]{Schroeder11}%
  \BibitemOpen
  \bibfield  {author} {\bibinfo {author} {\bibfnamefont {T.}~\bibnamefont
  {Schr\"oder}}, \bibinfo {author} {\bibfnamefont {F.}~\bibnamefont
  {G\"adeke}}, \bibinfo {author} {\bibfnamefont {M.~J.}\ \bibnamefont
  {Banholzer}}, \ and\ \bibinfo {author} {\bibfnamefont {O.}~\bibnamefont
  {Benson}},\ }\href {http://stacks.iop.org/1367-2630/13/i=5/a=055017}
  {\bibfield  {journal} {\bibinfo  {journal} {New J. Phys.}\ }\textbf {\bibinfo
  {volume} {13}},\ \bibinfo {pages} {055017} (\bibinfo {year}
  {2011})}\BibitemShut {NoStop}%
\bibitem [{\citenamefont {Doherty}\ \emph {et~al.}(2013)\citenamefont
  {Doherty}, \citenamefont {Manson}, \citenamefont {Delaney}, \citenamefont
  {Jelezko}, \citenamefont {Wrachtrup},\ and\ \citenamefont
  {Hollenberg}}]{Doherty13}%
  \BibitemOpen
  \bibfield  {author} {\bibinfo {author} {\bibfnamefont {M.~W.}\ \bibnamefont
  {Doherty}}, \bibinfo {author} {\bibfnamefont {N.~B.}\ \bibnamefont {Manson}},
  \bibinfo {author} {\bibfnamefont {P.}~\bibnamefont {Delaney}}, \bibinfo
  {author} {\bibfnamefont {F.}~\bibnamefont {Jelezko}}, \bibinfo {author}
  {\bibfnamefont {J.}~\bibnamefont {Wrachtrup}}, \ and\ \bibinfo {author}
  {\bibfnamefont {L.~C.}\ \bibnamefont {Hollenberg}},\ }\href {\doibase
  10.1016/j.physrep.2013.02.001} {\bibfield  {journal} {\bibinfo  {journal}
  {Physics Reports}\ ,\ \bibinfo {pages} {in press,}} (\bibinfo {year}
  {2013})},\ \bibinfo {note}
  {http://dx.doi.org/10.1016/j.physrep.2013.02.001}\BibitemShut {NoStop}%
\bibitem [{\citenamefont {Young}\ \emph {et~al.}(2013)\citenamefont {Young},
  \citenamefont {Hu},\ and\ \citenamefont {Rarity}}]{Young12}%
  \BibitemOpen
  \bibfield  {author} {\bibinfo {author} {\bibfnamefont {A.~B.}\ \bibnamefont
  {Young}}, \bibinfo {author} {\bibfnamefont {C.~Y.}\ \bibnamefont {Hu}}, \
  and\ \bibinfo {author} {\bibfnamefont {J.~G.}\ \bibnamefont {Rarity}},\
  }\href {\doibase 10.1103/PhysRevA.87.012332} {\bibfield  {journal} {\bibinfo
  {journal} {Phys. Rev. A}\ }\textbf {\bibinfo {volume} {87}},\ \bibinfo
  {pages} {012332} (\bibinfo {year} {2013})}\BibitemShut {NoStop}%
\bibitem [{\citenamefont {Bernien}\ \emph {et~al.}(2013)\citenamefont
  {Bernien}, \citenamefont {Hensen}, \citenamefont {Pfaff}, \citenamefont
  {Koolstra}, \citenamefont {Blok}, \citenamefont {Robledo}, \citenamefont
  {Taminiau}, \citenamefont {Markham}, \citenamefont {Twitchen}, \citenamefont
  {Childress},\ and\ \citenamefont {Hanson}}]{Bernien12}%
  \BibitemOpen
  \bibfield  {author} {\bibinfo {author} {\bibfnamefont {H.}~\bibnamefont
  {Bernien}}, \bibinfo {author} {\bibfnamefont {B.}~\bibnamefont {Hensen}},
  \bibinfo {author} {\bibfnamefont {W.}~\bibnamefont {Pfaff}}, \bibinfo
  {author} {\bibfnamefont {G.}~\bibnamefont {Koolstra}}, \bibinfo {author}
  {\bibfnamefont {M.~S.}\ \bibnamefont {Blok}}, \bibinfo {author}
  {\bibfnamefont {L.}~\bibnamefont {Robledo}}, \bibinfo {author} {\bibfnamefont
  {T.~H.}\ \bibnamefont {Taminiau}}, \bibinfo {author} {\bibfnamefont
  {M.}~\bibnamefont {Markham}}, \bibinfo {author} {\bibfnamefont {D.~J.}\
  \bibnamefont {Twitchen}}, \bibinfo {author} {\bibfnamefont {L.}~\bibnamefont
  {Childress}}, \ and\ \bibinfo {author} {\bibfnamefont {R.}~\bibnamefont
  {Hanson}},\ }\href {\doibase 10.1038/nature12016} {\bibfield  {journal}
  {\bibinfo  {journal} {Nature}\ }\textbf {\bibinfo {volume} {497}},\ \bibinfo
  {pages} {86} (\bibinfo {year} {2013})}\BibitemShut {NoStop}%
\end{thebibliography}

%

\end{document}